\documentclass[12pt]{article}
\usepackage[utf8]{inputenc}
\usepackage{setspace}
\usepackage{indentfirst}
\usepackage{amsthm,amsmath, amssymb}
\usepackage{natbib}
\usepackage{xcolor} 
\usepackage{graphicx} 
\usepackage{subfigure}
\usepackage{natbib}
\usepackage{hyperref}
\usepackage{bm}
\usepackage[top=2cm,left=3cm,bottom=2cm,right=3cm]{geometry}
\usepackage{url}
\usepackage{authblk}
\usepackage{multirow}

\usepackage{booktabs,rotating,subfigure,bm}
\usepackage{orcidlink}

\newtheorem{setting}{Setting}

\newtheorem{theorem}{Theorem}[section]

\def\BT{\mathbf{T}}
\def\BW{\mathbf{W}}
\def\BX{\mathbf{X}}
\def\Bx{\mathbf{x}}
\def\BZ{\mathbf{Z}}
\def\BL{\mathbf{L}}
\def\Bbeta{\boldsymbol{\beta}}
\def\Btheta{\boldsymbol{\theta}}
\def\Bzeta{\boldsymbol{\zeta}}

\def\Bell{\boldsymbol{\ell}}
\def\BSigma{\boldsymbol{\Sigma}}
\def\Bmeta{\boldsymbol{\eta}}
\def\Bpsi{\boldsymbol{\psi}}
\def\BD{\mathbf{D}}

\def\Be{\mathbf{e}}

\usepackage{xcolor}         

\definecolor{zhixin}{rgb}{0.6, 0.2, 0.8}
\definecolor{yanlin}{rgb}{0.8, 0.0, 0.0}

\begin{document}

\title{Heterogeneous Quantile Treatment Effect Estimation for Longitudinal Data with High-Dimensional Confounding}

\author{Zhixin Qiu$^{1}$, Huichen Zhu$^{2}$, Wenjie Wang$^{3}$ and Yanlin Tang$^{1,*}$\orcidlink{0000-0002-1950-903X}\thanks{Corresponding author. KLATASDS-MOE, School of Statistics, East China Normal University, Shanghai, 200062, China. Email: yltang@fem.ecnu.edu.cn}}
\affil[1]{\small KLATASDS-MOE, School of Statistics, East China Normal University, Shanghai, 200062, China}
\affil[2]{\small Department of Statistics, The Chinese University of Hong Kong, Hong Kong, 999077, China}
\affil[3]{\small Eli Lilly and Company, Indianapolis, Indiana, 46285, USA}


\maketitle

\begin{abstract}
{Causal inference plays a fundamental role in various real-world applications. However, in the motivating non-small cell lung cancer (NSCLC) study, it is challenging to estimate the treatment effect of chemotherapy on circulating tumor DNA (ctDNA). First, the heterogeneous treatment effects vary across patient subgroups defined by baseline characteristics. Second, there exists a broad set of demographic, clinical and molecular variables act as potential confounders. Third, ctDNA trajectories over time show heavy-tailed non-Gaussian behavior. Finally, repeated measurements within subjects introduce unknown correlation. Combining convolution-smoothed quantile regression and orthogonal random forest, we propose a framework to estimate heterogeneous quantile treatment effects in the presence of high-dimensional confounding, which not only captures effect heterogeneity across covariates, but also behaves robustly to nuisance parameter estimation error. We establish the theoretical properties of the proposed estimator. For statistical inference, we employ a Bootstrap of Little Bags procedure and establish its consistency in our setting. We further demonstrate its finite-sample performance through comprehensive simulations and illustrate its practical utility in the motivated NSCLC study.}
{Causal Inference; Double Machine Learning; Heterogeneous Quantile Treatment Effect; Longitudinal Data; Orthogonal Quantile Random Forest.}
\end{abstract}

\section{Introduction}	
Causal inference plays a fundamental role in a wide range of real-world applications, including personalized medicine, public policy evaluation, and the social and medical sciences. In recent years, growing attention has been devoted to the estimation of heterogeneous treatment effects, which allows researchers to understand how treatment effects vary across individuals or subgroups. In this paper, we consider a clinical application that highlights several key challenges in modern causal inference.

Our study is motivated by a real-world longitudinal dataset of patients with non-small cell lung cancer (NSCLC) that includes repeated measurements of circulating tumor DNA (ctDNA) as documented in the electronic health records. The richness and complexity of this data set pose several methodological challenges. First, the treatment effects are heterogeneous and vary across patient subgroups defined by baseline characteristics such as age, tumor stage, or genomic markers, motivating models that can capture the effect modification. Second, a broad set of demographic, clinical and molecular variables act as potential confounders, resulting in a high-dimensional confounding setting where conventional adjustment techniques may be inadequate or unstable. Third, as shown in Figure~\ref{fig:qq:ctdna}, ctDNA trajectories over time show heavy-tailed non-Gaussian behavior, likely due to biological heterogeneity and variable treatment responses; this undermines the reliability of mean-based estimators and highlights the need for a more robust estimate.  Finally, because each patient contributes multiple ctDNA measurements over time, the data set exhibits a longitudinal structure with potential within-subject correlation. 

Recently, heterogeneous causal effects have been widely studied. In particular, a variety of machine-learning methods have been proposed to model flexible treatment-response relationships and capture effect heterogeneity \citep{kunzel2019metalearners,bica2020estimating}. There is also a growing body of work on heterogeneous treatment effects under continuous treatments, extending causal inference tools to dose-response estimation \citep{kennedy2017non,doss2024nonparametric}. Among existing approaches, causal forests \citep{wager2018estimation,zhu2022hybrid} have become particularly popular for heterogeneous treatment effect estimation. Forest-based weights are more data-adaptive than classical kernel weights and can naturally handle mixed-type effect modifiers, and they also yield a relatively interpretable weighting scheme,which is often more convenient for statistical inference than many black-box machine-learning methods \citep{athey2019generalized}.

Modern applications often involve high-dimensional confounding variables, such as gene expressions, financial indicators, or wearable health sensor records. The presence of high-dimensional confounders poses a serious challenge to valid causal estimation, as standard methods may suffer from substantial regularization bias. The Double Machine Learning (DML) framework \citep{chernozhukov2018double,foster2023orthogonal} addresses this issue by constructing Neyman orthogonal score functions that are locally insensitive to estimation errors in nuisance parameters. 
Recently, \cite{pmlr-v97-oprescu19a,chen2025automaticdoublyrobustforests} combined random forests with DML techniques to estimate treatment effects in the presence of high-dimensional confounding.

Compared with ordinary least squares, quantile regression is more robust to outliers and heavy-tailed errors and it captures treatment effects across different quantiles. More recently, quantile treatment effects (QTE) have attracted growing attention.
For binary treatments, \cite{frolich2013unconditional} estimated unconditional QTE under endogeneity, \cite{giessing2023debiased} developed rank-score debiased inference for heterogeneous QTE with high-dimensional confounders and \cite{cheng2024doubly} studied doubly robust estimation for marginal structural quantile models. For continuous treatments, \cite{belloni2019valid} developed valid post-selection inference for high-dimensional quantile regression and \cite{baklicharov2024assumption} considered partially linear models for inference on conditional QTE. These works allow treatment effects to vary across quantile levels, but mainly focus on population-level effects. In contrast, our forest-based method accommodates covariate-dependent treatment effect heterogeneity.

Despite the advantages of quantile regression, its integration with machine learning methods—particularly GRF—is technically challenging because the quantile score is nondifferentiable. One prominent approach is to replace the check loss with a convolution-smoothed quantile loss \citep{fernandes2021smoothing,tan2022high}, which is a twice continuously differentiable convex approximation. This approximation enables the use of gradient-based optimization and facilitates integration with GRF. Standard QRF uses mean-based CART splitting and only incorporates quantiles at the prediction stage. By using a quantile-based splitting criterion, our method is more robust to heavy-tailed errors, as shown in Section S2.2 of the Supplementary Materials.

In this paper, our aim is to estimate the heterogeneous quantile treatment effect in the presence of high-dimensional confounding, while ensuring robustness to heavy-tailed noise in the response. Our main contributions are as follows. We propose a forest-based local quantile regression to obtain a nonparametric quantile treatment effect estimation, designed to be robust against heavy-tailed errors. To address high-dimensional confounding, we introduce the Orthogonal Quantile Random Forest (OQRF), which incorporates Neyman orthogonality into the splitting criterion to reduce bias from nuisance parameter estimation. For theoretical development, we address a critical issue in the proof by using a convolution-smoothed quantile loss function. The convolution smoothing allows us to use gradient-based methods in the proof process, which enables us to obtain a tighter error bound for the parameter of interest.

The rest of the paper is organized as follows. In Section~\ref{sec:method}, we introduce the forest-based nonparametric estimation method for quantile treatment effects and describe the Bootstrap of Little Bags procedure for statistical inference. In Section~\ref{sec:theoretical_properties}, we present the theoretical results, including the error bound, the asymptotic normality of the proposed estimator and the asymptotic validity of the Bootstrap of Little Bags confidence interval. In Section~\ref{sec:sim}, we present simulation studies to compare the proposed method with other existing approaches. In Section~\ref{sec:appl}, we apply the proposed method to motivated non-small cell lung cancer data.

\section{Methods}\label{sec:method}
\subsection{Model}

We consider a longitudinal study of $2n$ subjects, where each subject $i$ contributes $m_i$ repeated measurements denoted by $\{Y_{ij}, \BT_{ij}, \BW_{ij}\}_{j=1}^{m_i}$ and a vector of baseline modifiers $\BX_i$. 
Here, $Y_{ij}\in\mathbb{R}$ denotes the response (say, clinical outcome reflecting the severity of the disease); $\BT_{ij}\in\mathbb{R}^{p_t}$ are treatment variables whose effects we aim to estimate (say, doses of investigational drugs); $\BW_{ij}\in\mathbb{R}^{p_w}$ are high-dimensional confounders whose dimension $p_w$ grows with the sample size (say, dynamic biomarkers such as blood pressure, heart rate, laboratory measurements); and $\BX_i\in\mathbb{R}^{p_x}$ are subject-level covariates that induce heterogeneity in both treatment and confounder effects (say, baseline demographics such as age, weight, gender).
To model heterogeneous effects at the quantile level $\tau$, we posit the following conditional quantile function,
\begin{align}\label{eq:main_model}
Q_\tau\left(Y_{ij} \mid \mathbf{T}_{ij}, \mathbf{W}_{ij}, \mathbf{X}_i\right) &= \boldsymbol{\theta}^{\star}_\tau(\mathbf{X}_i)^{\top} \mathbf{T}_{ij} + \boldsymbol{\beta}^{\star}_\tau(\mathbf{X}_i)^{\top} \mathbf{W}_{ij}.
\end{align}
The vector‐valued function \(\boldsymbol\theta^{\star}_\tau(\mathbf X_i)\) captures how the treatment effect at quantile \(\tau\) varies with baseline modifiers, while \(\boldsymbol\beta^{\star}_\tau(\mathbf X_i)\) encodes confounder effects.
Without loss of generality, we include a constant term in $\BW_{ij}$, ensuring that $\Bbeta_\tau^{\star}(\BX_i)$ contains an intercept.

Since $\BW_{ij}$ are confounders, which influences both $Y_{ij}$ and $\BT_{ij}$, we further assume that the treatment can be linearly projected onto the confounders,
\begin{align}\label{eq:T_model*}
    \mathbf{T}_{ij} &= \BL^{\star}(\BX_i)^{\top} \BW_{ij} + \Be_{ij}, 
\end{align}
where $\BL^{\star}(\BX_i)\in\mathbb{R}^{p_w\times p_t}$. Equation \eqref{eq:T_model*} divides $\BT_{ij}$ into the systematic component $\BL^{\star}(\BX_i)^{\top} \BW_{ij}$ explained by confounders and the residual $\Be_{ij}$ capturing exogenous fluctuations uncorrelated with $\BW_{ij}$ given $\BX_i$. Finally, we assume a location-shift error structure
$$\varepsilon_{ij,\tau} = Y_{ij} - \boldsymbol{\theta}^{\star}_\tau(\mathbf{X}_i)^{\top} \mathbf{T}_{ij} - \boldsymbol{\beta}^{\star}_\tau(\mathbf{X}_i)^{\top} \mathbf{W}_{ij},$$
where $\varepsilon_{ij,\tau}$ are marginally identically distributed across all $(i,j)$ and independent across subjects, but may exhibit arbitrary unknown dependence within each subject, and $\varepsilon_{ij,\tau}$ is assumed to be independent of the treatment noise $\mathbf{e}_{ij}$.
Under this setup, only the intercept of $\Bbeta^{\star}_\tau(\BX_i)$ varies with $\tau$. For notation ease, we hereafter suppress the subscript $\tau$ from $\boldsymbol{\theta}_\tau$.

Our causal interpretation relies on the structural quantile model \eqref{eq:main_model}. The main focus $\Btheta_{\tau}^{\star}(\BX_i)$ represents a stage-specific structural conditional quantile effect of the current treatment, rather than a marginal effect of an entire longitudinal treatment regime, which is aligned with recent machine-learning approaches for panel data that target a contemporaneous treatment-effect parameter \citep{emmenegger2023plug,fuhr2024double}. Such attention is clinically useful because treatment effects often changes over time with patients' biological status.

The model (\ref{eq:main_model}) is driven by two key needs in precision medicine: (i) the ability to adjust for rich, high‐dimensional, time‐varying confounders when estimating causal treatment–response relationships, and (ii) the desire to capture heterogeneous effects across the entire outcome distribution rather than just the mean.   
The quantile regression formulation in \eqref{eq:main_model} then yields dose–response curves that vary flexibly with $\BX_i$, enabling tailored treatment recommendations that account for each patient’s unique biomarker trajectory and demographic profile.


\subsection{Heterogeneous Quantile Treatment Estimation}\label{sec:theta_est}

Equation \eqref{eq:main_model} specifies a linear quantile regression model in which the treatment effect $\Btheta^{\star}(\Bx)$ varies with the effect modifiers $\Bx$. Estimation is challenging for three reasons. First, treatment-effect heterogeneity must be captured flexibly across $\Bx$. Second, high-dimensional confounders $\BW$ require regularization, and regularization or overfitting in the nuisance components may induce bias in $\Btheta^{\star}(\Bx_0)$. Third, subjects are observed repeatedly at irregular time points, which makes standard quadratic inference function (QIF) methods that rely on aligned observation schedules inapplicable \citep{qu2006quadratic}. Our estimation strategy addresses these three difficulties through complementary components.
Our estimation strategy tackles these challenges through three complementary components. 
First, we deploy a forest-based estimator to flexibly model effect heterogeneity across the covariate space. Compared with kernel weights, forest-based weights are more adaptive, automatically selecting the effective neighborhood around $\Bx_0$ without requiring bandwidth specification; also, they are naturally suited to mixed-type effect modifiers, including continuous, categorical and binary variables because the similarity is defined through shared local partitions rather than Euclidean distance \citep{athey2019generalized}.
Second, we construct a Neyman-orthogonal score that attenuates bias arising from high-dimensional nuisance parameters. 
Third, we implement a subject-level downweighting rule to equalize the influence of each participant, regardless of the number of observations \citep{datta2014robust}.
Together, these elements deliver an estimator that recovers treatment heterogeneity across both covariates and the full quantile distribution of the outcome.
We describe the estimation procedure in detail below.

{\it Orthogonal Quantile Score Function.} To obtain a debiased QTE estimator, we build an orthogonal quantile score framework \citep{belloni2019valid}. In the presence of high-dimensional confounding, we employ a Neyman orthogonal quantile score function to eliminate the bias introduced by the estimation of nuisance parameter. Specifically, Neyman orthogonality means that the first-order derivative of the expected score with respect to the nuisance parameter is zero at the true nuisance value. This condition ensures that small errors in the estimation of nuisance parameters have only a second-order effect on the estimation of the treatment effect. We define the orthogonal score function as
\begin{equation}\label{orthogonal}
\Bpsi_\tau(Y,\BT,\BW,\BX;\Btheta,\Bbeta,\BL)
= \varphi_\tau\bigl\{Y - \Btheta(\BX)^\top \BT - \Bbeta(\BX)^\top \BW\bigr\}\;\bigl\{\BT - \BL(\BX)^\top \BW\bigr\},
\end{equation}
where $\varphi_\tau(u)=\tau-I\left(u\leq 0\right)$ is the subgradient of the check loss. Denote $\Bmeta=(\Bbeta,\BL)$ as the nuisance parameter. See Section S5.1 in the online Supplementary Materials for the proof of the Neyman orthogonality.

Having introduced the orthogonal score and its key property, we now present a brief outline of our method. To simplify the exposition, we first assume that the forest is already built. The detailed procedure for building the forest will be introduced later in Section~\ref{sec:forest_weight}. 

\noindent Step 0. We split the data set $\mathcal{D}$ equally into two disjoint subsets, $\mathcal{D}_1=\left\lbrace 1,\dots,n\right\rbrace$ and $\mathcal{D}_2=\left\lbrace n+1,\dots,2n\right\rbrace$, according to the subject indices, where $\mathcal{D}_1$ is used to estimate nuisance parameters and $\mathcal{D}_2$ is for the parameter of interest.

\noindent \textbf{Step 1}. We construct two separate random forests to assign each subject a similarity weight $\alpha_i(\Bx_0)$, capturing its relevance to the target point $\Bx_0$ and facilitating localized estimation. The first forest is built using $\mathcal{D}_1$ and assigns weights to subjects in $\mathcal{D}_1$, while the second forest is built using $\mathcal{D}_2$ and assigns weights to subjects in $\mathcal{D}_2$.

\noindent \textbf{Step 2}. Estimate $\mathbf{L}^{\star}(\Bx_0)$ and $\Bbeta^{\star}_\tau(\Bx_0)$ with $\mathcal{D}_1$, denoted by $\tilde{\mathbf{L}}(\Bx_0)$ and $\tilde{\Bbeta}_\tau(\Bx_0)$. 

\noindent \textbf{Step 3}. Estimate $\Btheta^{\star}(\Bx_0)$ with $\mathcal{D}_2$ by solving the Neyman orthogonal estimating equation using plug-in estimators $\tilde{\mathbf{L}}(\Bx_0)$ and $\tilde{\Bbeta}_\tau(\Bx_0)$, denoted by $\hat{\Btheta}(\Bx_0)$.

In Step 1, we get the similarity weights $\alpha_i(\Bx_0)$ via random forest \citep{athey2019generalized},
$$\alpha_{ib}(\Bx_0)=\frac{1\{\BX_i\in \mathcal{L}_b(\Bx_0)\}}{\left|\mathcal{L}_b(\Bx_0)\right|},\quad\alpha_i(\Bx_0)=\frac{1}{B}\sum_{b=1}^B\alpha_{ib}(\Bx_0),$$
where $\mathcal{L}_b(\Bx_0)$ is the leaf of $b$-th tree that contains $\Bx_0$, and $B$ is the number of trees. 

In Step 2, we estimate $\mathbf{L}^{\star}(\Bx_0)$ using a weighted Lasso regression of $\BT$ on $\BW$ and estimate $\Bbeta^{\star}_\tau(\Bx_0)$ using a weighted $\ell_1$-QR of $Y$ on $\BT$ and $\BW$; see Section S1.1 in Supplementary Materials for details.

In Step 3, we propose to estimate $\Btheta^{\star}(\Bx_0)$ by solving \eqref{orthogonal} with a forest-based weight $\alpha_i(\Bx_0)$ and plugged-in nuisance estimator,
  \begin{equation}\label{eq:estimating equation} 
  \sum_{i\in\mathcal{D}_2}\alpha_i(\Bx_0)\sum_{j=1}^{m_i}\frac{1}{m_i}\varphi_\tau\left\{Y_{ij}-\Btheta^{\top}\BT_{ij}-\tilde{\Bbeta}_\tau(\Bx_0)^{\top}\BW_{ij}\right\}\left\{\BT_{ij}-\tilde{\BL}(\Bx_0)^{\top}\BW_{ij}\right\}=\bm{0}.
  \end{equation}
A down-weighting scheme is applied to ensure that each subject contributes equally to the estimation, preventing subjects with more measurements from having a disproportionate influence on the estimator. Since we assume that $m_i$ is finite, the down-weighting scheme will not affect the statistical efficiency. Since equation~\eqref{eq:estimating equation} may not yield an exact zero, we define $\hat{\Btheta}(\Bx_0)$ as the solution to the following minimization problem,
  \begin{equation}\label{eq:objective}
        \mathop{\arg\min}_{\Btheta\in\Theta}\left\| \sum_{i\in\mathcal{D}_2}\alpha_i(\Bx_0)\sum_{j=1}^{m_i}\frac{1}{m_i}\varphi_\tau\left\{Y_{ij}-\Btheta^{\top}\BT_{ij}-\tilde{\Bbeta}_\tau(\Bx_0)^{\top}\BW_{ij}\right\}\left\{\BT_{ij}-\tilde{\BL}(\Bx_0)^{\top}\BW_{ij}\right\}\right\|.
  \end{equation}
The iterative algorithm used to solve \eqref{eq:objective} is described in Section~S1.2 of the online Supplementary Materials.
  
\subsection{Orthogonal Quantile Forest Construction}\label{sec:forest_weight}

In this subsection, we describe the construction of Orthogonal Quantile Forest, treating each subject as a unit. Let $\hat{\Btheta}_P$ denote the local estimator in the parent node, and $\hat{\Btheta}_{C_1}$, $\hat{\Btheta}_{C_2}$ be the corresponding estimators in the child nodes. Our goal is to find a split that maximizes the discrepancy between $\hat{\Btheta}_{C_1}$ and $\hat{\Btheta}_{C_2}$, improving homogeneity within each leaf. However, calculating $\hat{\Btheta}_{C_1}$ and $\hat{\Btheta}_{C_2}$ for any candidate splitting is time-consuming. To alleviate this, we use one-step Newton updates to approximate $\hat{\Btheta}_{C_1}$ and $\hat{\Btheta}_{C_2}$, based on the gradient and Hessian of a convolution-smoothed quantile loss \citep{tan2022high}. 

{\it Splitting criterion.} \textbf{Step (i). Estimate the nuisance parameter $\left( \tilde{\BL}_P,\tilde{\Bbeta}_P\right)$.}
 The $\nu$-th column of $\tilde{\BL}_P$ can be estimated by
$$\tilde{\Bell}_P^{(\nu)}\in\mathop{\arg\min}_{\Bell^{(\nu)}}\left\lbrace\frac{1}{n_P}\sum_{\{i \in P\}}\frac{1}{m_i}\sum_{j=1}^{m_i}\left\{T_{ij}^{(\nu)}-\Bell^{(\nu){\top}}\BW_{ij}\right\}^2+\frac{\lambda_{1}^{(\nu)}}{n_P}\|\Bell^{(\nu)}\|_1\right\rbrace,$$
Denote $\BD_{ij}=(\BT_{ij}^{\top},\BW_{ij}^{\top})^{\top}$, $\Bzeta_\tau=(\Btheta^{\top},\Bbeta_\tau^{\top})^{\top}$. Then $\Bbeta_P$ can be estimated by
$$ \tilde{\Bzeta}_P=\left( \tilde{\Btheta}_P,\tilde{\Bbeta}_P\right) \in\mathop{\arg\min}_{\Bzeta}\left\lbrace\frac{1}{n_P}\sum_{\{i\in P\}}\frac{1}{m_i}\sum_{j=1}^{m_i}\rho_{\tau h}(Y_{ij}-\Bzeta^{\top}\BD_{ij})+\frac{\lambda_{2}}{n_P}\|\Bzeta\|_1\right\rbrace,$$
where $\rho_{\tau h}(\cdot)$ denotes the convolution-type smoothed quantile loss defined as
$$\rho_{\tau h}\left(Y_{ij}-\Bzeta^{\top}\BD_{ij}\right):=\int_{-\infty}^\infty \rho_\tau\big(u\big)\cdot K_h\left\{u-(Y_{ij}-\Bzeta^{\top}\BD_{ij})\right\} du,$$ 
with $K_h(t)=(1/h)K(t/h)$; $K(\cdot)$ is a kernel function and $h$ is the bandwidth. 
   
\noindent \textbf{Step (ii). Estimate $\Btheta_P$.}
	$\hat{\Btheta}_P\in\arg\min_{\Btheta}\left\|\frac{1}{n_P}\sum_{\{i\in P\}}\frac{1}{m_i}\sum_{j=1}^{m_i}\psi_\tau(\BZ_{ij};\Btheta,\tilde{\Bmeta}_P)\right\|.$
    
\noindent \textbf{Step (iii). Calculate the Hessian matrix $\mathbf{A}_P$.}
We adopt a convolution-type smoothed quantile score function. Decompose the treatment variable $\BT_{ij}$ into $\tilde{\BL}_P^{\top}\BW_{ij}$ and the residual $\tilde{\Be}_{ij}=\BT_{ij}-\tilde{\BL}_P^{\top}\BW_{ij}$, then the smoothed quantile loss is given by
\begin{equation}
    Q_{\tau h}(\Btheta,\tilde{\Bmeta})=\frac{1}{n_P}\sum_{\{i\in P\}}\frac{1}{m_i}\sum_{j=1}^{m_i}\int_{-\infty}^\infty \rho_\tau\big(u\big)\cdot K_h\left\{ u-(\tilde{Y}_{ij}-\Btheta^{\top}\tilde{\Be}_{ij})\right\} du,
    \nonumber
\end{equation}
 where $\tilde{Y}_{ij}=Y_{ij}-\tilde{\Bbeta}^{\top}\BW_{ij}-\Btheta^{\top}\tilde{\BL}_P^{\top}\BW_{ij}$. Let $\bar{K}(u) = \int_{-\infty}^u K(t)dt$ denote the integrated kernel function. The score function induced by $Q_{\tau h}(\Btheta,\tilde{\Bmeta})$ is 
\begin{equation}
	\begin{aligned}
	&\quad\frac{1}{n_P}\sum_{\{i\in P\}}\frac{1}{m_i}\sum_{j=1}^{m_i} \left[\tau-\bar{K}
    \left\{(\Btheta^{\top}\tilde{\Be}_{ij}-\tilde{Y}_{ij})/h\right\}\right] \tilde{\Be}_{ij}\\
        &=\frac{1}{n_P}\sum_{\{i\in P\}}\frac{1}{m_i}\sum_{j=1}^{m_i}\left[\tau-\bar{K}\left\{(\Btheta^{\top}\BT_{ij}+\tilde{\Bbeta}_P^{\top}\BW_{ij}-Y_{ij})/h\right\}\right]\tilde{\Be}_{ij}.
	\end{aligned}
    \nonumber
\end{equation}
Then compute the local Hessian matrix 
\begin{equation}
    \mathbf{A}_P:=\frac{1}{n_P}\sum_{\{i\in P\}}\frac{1}{m_i}\sum_{j=1}^{m_i}-K_h\left(\Btheta^{\top}\BT_{ij}+\tilde{\Bbeta}_P^{\top}\BW_{ij}-Y_{ij}\right)\tilde{\Be}_{ij}\BT_{ij}^{\top}.
    \nonumber
\end{equation}
  
\noindent \textbf{Step (iv). Split $P$ into $C_1$ and $C_2$ via maximizing the heterogeneity score.}
We first define the subject-level influence score as $\boldsymbol{\rho}_i=\mathbf{A}_P^{-1}\left\{\frac{1}{m_i}\sum_{j=1}^{m_i}\psi_\tau\left(\BZ_{ij};\hat{\Btheta}_P,\tilde{\Bmeta}_P\right)\right\}$. Let $\rho_{i}^{(\nu)}$ be the $\nu$-th component of $\boldsymbol{\rho}_i$. For $\nu=1,2,\dots,q$, define
$\tilde{\Delta}_\nu(C_1,C_2)=\sum_{j=1}^2\frac{1}{n_{C_j}}\left\{ \sum_{\{i\in C_j\}}\rho_{i}^{(\nu)}\right\}^2$,
The final heterogeneity score is
$$\Delta^*(C_1,C_2):=\mu\max_\nu\tilde{\Delta}_\nu(C_1,C_2)+\left(1-\mu \right)\frac{1}{q}\sum_{\nu=1}^{q}\tilde{\Delta}_\nu(C_1,C_2),$$
where $\mu\sim\text{Uniform}(0,1)$. This convex combination of the maximum and average $\tilde{\Delta}_\nu$ values guides the split of a parent node into two children $C_1$ and $C_2$ by maximizing $\Delta^*(C_1, C_2)$.

\subsection{Bootstrap of Little Bags}
We use the bootstrap of little bags (BLB) to estimate the variance of the proposed estimator. The key difficulty is that the forest weights are adaptive and depend on random subsampling and tree construction, making a direct analytic variance estimator impractical. BLB provides a computationally efficient way to approximate the variance for forest-type estimators.

Specifically, we randomly split $\mathcal{D}_2$ into $G$ little bags. For each bag $g=1,\ldots,G$, we grow $R$ trees using subsamples drawn from that bag and compute the corresponding tree-level estimating score at the target point $\Bx_0$, denoted by $\bm{\Psi}_{g,r}(\Bx_0)$, $r=1,\ldots,R$.
Let
\[
\bar{\bm{\Psi}}_g(\Bx_0)=R^{-1}\sum_{r=1}^R \bm{\Psi}_{g,r}(\Bx_0),
\qquad
\bar{\bm{\Psi}}(\Bx_0)=G^{-1}\sum_{g=1}^G\bar{\bm{\Psi}}_g(\Bx_0).
\]
The between-bag variation captures sampling variability, while the within-bag variation corrects for the Monte Carlo noise induced by tree construction. Following the ANOVA decomposition of BLB, we estimate the score variance by 
$
\hat{\bm H}_n(\Bx_0)
=
\hat{\bm V}_{\mathrm{total}}(\Bx_0)
-
\hat{\bm V}_{\mathrm{within}}(\Bx_0)$,
where $\hat{\bm V}_{\mathrm{total}}(\Bx_0)$ is the sample covariance of $\{\bar{\bm{\Psi}}_g(\Bx_0)\}_{g=1}^G$ and $\hat{\bm V}_{\mathrm{within}}(\Bx_0)$ is the average within-bag covariance over the $R$ trees.
Combining this BLB estimator with a plug-in estimator $\hat{\bm M}(\Bx_0)$ of the local Jacobian gives the sandwich variance estimator
$\hat{\bm\Sigma}_n^2(\Bx_0)
=
\hat{\bm M}(\Bx_0)^{-1}
\hat{\bm H}_n(\Bx_0)
\left\{\hat{\bm M}(\Bx_0)^{-1}\right\}^{\top}.$
With this variance estimator, we can construct the corresponding confidence intervals for QTE. The rationale behind this construction and further implementation details are provided in Section S1.3 of the Supplementary Materials.

\section{Theoretical Properties}\label{sec:theoretical_properties}
In this section, we present theoretical results for the estimation of the treatment effect. The technical assumptions required are listed in Section S4.1 of the online Supplementary Materials, with a rigorous discussion. With these preparatory results in Section S4.2, we now establish the main theoretical guarantees for the treatment effect estimator. Theorems~\ref{Consistency}-\ref{Asymptotic Normality} show the consistency, error bound, and asymptotic normality of $\hat{\Btheta}(\Bx_0)$.
 \begin{theorem}{\textbf{(Consistency)}}\label{Consistency}
Under Assumptions 1-9 given in the online Supplementary Materials, let $s$ denote the subsample size and $B$ denote the number of trees. Assume that $B\geq\frac{n}{s}$, with $s=o(n)$ and $s\rightarrow\infty$ as $n\rightarrow\infty$, then $\left\| \hat{\Btheta}(\Bx_0)-\Btheta^{\star}(\Bx_0)\right\| =o_p(1)$.
\end{theorem}
	
\begin{theorem}{\textbf{(Error Bound)}}\label{Error Bound}
	Under Assumptions 1-10 in the online Supplementary Materials, $\mathbb{E}\left\{ \left\|\hat{\Btheta}(\Bx_0)-\Btheta^{\star}(\Bx_0)\right\| \right\}= O\left( s^{-\frac{1}{2\omega p_x}}+\sqrt{\frac{s}{n}}\right)$, where $\omega$ is defined in Assumption 10.
\end{theorem}

\begin{theorem}{\textbf{(Asymptotic Normality)}}\label{Asymptotic Normality}
Under Assumptions 1-10 given in the online Supplementary Materials, further assume the subsample size $s=O(n^b)$ for some $b\in\left(1-\frac{1}{1+\omega p_x},1\right)$, with $\omega$ defined in Assumption 10. For any vector $\bm{a}\in\mathbb{R}^{p_t}$, with $\|\bm{a}\|= 1$, there exists a sequence $\sigma_n(\Bx_0,\bm{a})$ that satisfies $\sigma_n(\Bx_0,\bm{a})=O\left(\sqrt{\mathop{polylog}(n/s)^{-1}s/n}\right)$,
\begin{equation}\label{AN}
	\sigma_n(\Bx_0,\bm{a})^{-1}\left\langle \bm{a},\hat{\Btheta}(\Bx_0)-\Btheta^{\star}(\Bx_0)\right\rangle\overset{d}{\longrightarrow}\mathcal{N}(0,1).
\end{equation}
Here, $polylog(n/s)$ denotes a positive function that is bounded away from zero and grows at most polynomially in $\log(n/s)$.
\end{theorem}

As shown in \eqref{AN}, we obtain the same convergence rate in \cite{athey2019generalized} as for the finite-dimensional confounding setting. \cite{pmlr-v97-oprescu19a} obtained the same convergence rate for mean regression in the presence of high-dimensional confounders; however, their theoretical guarantees no longer hold when the variance of the error term goes to infinity.

We next justify Bootstrap of Little Bags inference for the proposed estimator. Under standard conditions, the resulting bootstrap variance estimator is consistent.
\begin{theorem}\label{BLB validity}
Under assumptions in Theorem~\ref{Asymptotic Normality}, 
$\left\|\hat{\bm H}_n(\Bx_0)-\bm H_n(\Bx_0)\right\|_{F}/\left\|\bm H_n(\Bx_0)\right\|_{F}\overset{p}{\longrightarrow}0$. Define the plug-in covariance estimator
$\hat{\bm \Sigma}^2_n(\Bx_0)=\left\{\hat{\bm M}(\Bx_0)^{-1}\right\}\hat{\bm H}_n(\Bx_0)\left\{\hat{\bm M}(\Bx_0)^{-1}\right\}^{\top}.$
Then for each $\nu=1,\dots,p_t$ and any given $\alpha\in(0,1)$, the Wald-type confidence interval
$\hat{\Btheta}^{(\nu)}(\Bx_0)
\;\pm\;z_{1-\alpha/2}\sqrt{\bm e_\nu^\top\hat{\bm \Sigma}^2_n(\Bx_0)\bm e_\nu}$
has asymptotic coverage $1-\alpha$.
\end{theorem}
Theorem~\ref{BLB validity} shows the bootstrap variance estimator is asymptotically valid and thus justifies the resulting confidence intervals achieve the asymptotic coverage.

\section{Simulation Studies}\label{sec:sim}
In this section, we evaluate the empirical performance of our proposed estimator against a suite of established alternatives. 

\begin{itemize}

    \item \textbf{Orthogonal Random Forest \citep[ORF,][]{pmlr-v97-oprescu19a}}. ORF targets mean treatment effects using a least squared loss, with orthogonal score derived from residual-on-residual regression.

    \item \textbf{Double Machine Learning with Lasso \citep[DML-Lasso,][]{chernozhukov2018double}}. The heterogeneity is addressed by creating an expanded linear base of parameters. Nuisance functions are estimated using polynomial regression with a Lasso penalty, then be plugged into a second-stage polynomial regression to estimate the treatment effect.

    \item \textbf{Double Machine Learning with Random Forest (DML-RF)}. Both nuisance parameters and treatment effects are estimated by the random forest.

    \item \textbf{Double Machine Learning with Quantile Lasso \citep[DML-QR,][]{belloni2019valid}}. DML-QR extends the DML idea to quantile treatment effect estimation.  The heterogeneity is addressed by a polynomial spline basis expansion, same as DML-Lasso.

    \item \textbf{Generalized Random Forest \citep[GRF,][]{athey2019generalized}}. We use a quantile-score-based version of GRF target to the heterogeneous quantile treatment effects.
\end{itemize}

We include ORF as a mean-based forest benchmark to highlight the robustness of our method under heavy-tailed errors. DML-Lasso and DML-RF are included as orthogonalization-based baselines with parametric sparse and nonparametric nuisance estimation, respectively. We further include two quantile-based competitors, GRF and DML-QR, to compare OQRF with methods that also target QTE.

\subsection{Simulation Settings}
We assess performance under the following data‐generating process.  
For each subject $i = 1,\dots,2n$ with $n=1000$, and each observation $j = 1,\dots,m_i$ with $m_i$ uniformly from $\{3,4,5,6\}$, we simulate data from the following model
\begin{equation*}
    Y_{ij}=\theta^{\star}(\BX_i)T_{ij} +\Bbeta^{\star}(\BX_i)^{\top}\BW_{ij}+\varepsilon_{ij},\qquad T_{ij}=\Bell^{\star}(\BX_i)^{\top}\BW_{ij}+e_{ij},
\end{equation*}
where $\BW_{ij,1}=1$, $\BW_{ij,-1}\sim N(0,\BSigma_\BW)$, and the $(p,q)$-th component of $\BSigma_\BW$ is $\sigma_{pq}=0.5^{|p-q|}$. So that the first entry of $\BW_{ij}$ is an intercept and the remaining $p_W-1$ features follow a mean-zero Gaussian with AR(1) covariance. The treatment noise $e_{ij}\sim\mathrm{Unif}(-1,1)$.     
We consider two high‐dimensional settings with $\dim(\mathbf W_{ij}) = 201$ and $501$ respectively, keeping the sparsity level at $k=5$.
We describe other ata generating process details as follows.

\begin{setting}\label{constant_nuisance}
In this setting, we consider a one-dimensional effect modifier and fixed nuisance parameters. The effect modifiers $X_i$ are drawn i.i.d. from $\mathrm{Unif}(0,1)$. The nuisance parameters are set as $$ \Bbeta^{\star}(x)=(0,0.5_5^{\top},0,\dots,0),$$
$$\Bell^{\star}(x)=(0_6^{\top},0.5,0.5,-0.5,-0.5,-0.5,0,\dots,0)$$.
The treatment effect $\theta^{\star}(x)$ is defined as a piecewise linear function
\begin{equation}\label{treatment setting}
	\theta^{\star}(x)=\left\{
	\begin{array}{cl}
		1+0.5x &  0\leq x <0.3,\\
		1.15+3(x-0.3) &  0.3\leq x <0.6, \\
		2.05-1.5(x-0.6)& 0.6\leq x \leq1. \\
	\end{array} \right.
\end{equation}
\end{setting}

\begin{setting}\label{vc_nuisance}
This setting extends Setting \ref{constant_nuisance} by allowing the nuisance parameters to vary with $x$. In this case, the modifiers $X_i$ and the treatment effect remain the same as those in Setting \ref{constant_nuisance}. The nuisance parameters are defined as $$\Bbeta^{\star}(x)=\left(0,\frac{x+3}{6},\frac{\sin(\pi x)}{2},(1-x)^2,\frac{1}{2},\frac{1}{2},0,\dots,0\right)$$ and $$\Bell^{\star}(x)=\left(0_6^{\top},\frac{x^3+1}{4},\frac{\cos\left((6x-5)\pi/3\right)}{2},\frac{1}{2+2x},-\frac{1}{2},-\frac{1}{2},0,\dots,0\right).$$
\end{setting}

\begin{setting}\label{2-dim modifier}
In this setting, we consider a two-dimensional effect modifier $\BX_i=\left(X_{i,1},X_{i,2}\right)$ with $X_{i,1}\sim \mathrm{Unif}(0,1),X_{i,1}\sim Bern(0.5)$. The nuisance parameters $\Bbeta^{\star}(\Bx)$ and $\Bell^{\star}(\Bx)$ are specified as in Setting \ref{constant_nuisance}. The treatment effect is defined as $\theta^{\star}(\Bx)=\theta_1^{\star}\left(x_1\right)\cdot I\left(x_2=0\right)+\theta_2^{\star}
\left(x_1\right)\cdot I\left(x_2=1\right),$
where $\theta^{\star}_1(x)$ is defined as in equation \eqref{treatment setting} and $\theta^{\star}_2(x)$ is defined as follows: 
\begin{equation}
	\theta^{\star}_2(x)=\left\{
	\begin{array}{cl}
		1.5x^2+0.5&  0\leq x <0.2,\\
		2x^2+4x-0.32 &  0.2\leq x <0.6, \\
		0.5x+2.5 & 0.6\leq x \leq1.\\
	\end{array} \right.
	\nonumber
\end{equation}

\end{setting}
To assess robustness to noise, we generate the error term $\varepsilon_{ij}$ from three different distributions in each three settings: (i) normal distribution, (ii) $t$-distribution with 3 degrees of freedom, and (iii) Cauchy distribution. These distributions represent increasing levels of heavy-tailed. Furthermore, to reflect the dependence of the subject within the subject, we introduce a weak correlation between repeated measures: for each subject $i$, the error vector $\boldsymbol{\varepsilon}_i=(\varepsilon_{i1},\dots,\varepsilon_{im_i})^\top$ is drawn from a multivariate version of the corresponding distribution with the correlation matrix $\BSigma_\varepsilon$, where the entry $(p,q)$-th is given by $\sigma_{pq} = 0.5^{|p - q|}$.

\subsection{Results}
For each method, we compute two performance metrics: the bias and the mean integrated squared error (MISE) of the estimated treatment effect.
The numerical results are shown in Table~\ref{tab:1}, where the reported values are Monte Carlo averages over $500$ replications and we present the corresponding Monte Carlo standard errors in parentheses.

For OQRF, we set the tree number $B=500$, subsample ratio $s/n=0.5$, the max tree depth of $15$ and the minimum leaf size of $20$. 
We take the bandwidth as $h=\max\left\lbrace\frac{\sqrt{\tau(1-\tau)}}{3}\left\{\frac{s\log (p_T+p_W)}{n}\right\}^{1/4},0.1\right\rbrace$. The first penalty level is set via $\frac{\lambda_1}{n_{rf}}=\frac{c}{100}\sqrt{\frac{s\log p_W}{n}}$, where $c$ is selected from $\{1,2,\dots,10\}$ using BIC. The second penalty level $\lambda_2$ is selected following the procedure described in S1.1 of the online Supplementary Materials.
To adapt to the longitudinal data setting, we also apply a down-weighting scheme across all comparison methods. For ORF, the hyperparameter follows \cite{pmlr-v97-oprescu19a} with $B=200$, subsample size $s=\left(\frac{n}{\log p_W}\right)^{0.88}$, the max tree depth of $20$ and the minimum leaf size of $5$. Both $\frac{\lambda_1}{n_{rf}}$ and $\frac{\lambda_2}{n_{rf}}$ are set to be $\sqrt{\frac{\log p_W}{10n}}$. For DML-Lasso, we use a polynomial basis of degree $3$, and the penalty term is selected via cross-validation. For DML-RF, both nuisance parameters and treatment effect are estimated by a random forest regressor with $B=100$ trees, maximum tree depth of $20$ and the minimum leaf size of $5$.


Under standard normal errors, the proposed OQRF performs comparably to the baseline methods. Although ORF and DML-Lasso achieve slightly smaller bias in certain settings, OQRF still yields smaller MISE, indicating better overall estimation accuracy. 
ORF does not always achieve the smallest MISE even under normal errors, possibly because we follow the default tuning choices in \citet{pmlr-v97-oprescu19a}, such as the maximum number of splits and the minimum leaf size, which may not be optimal for the present setting. 
As the error distribution becomes heavier-tailed, the advantage of OQRF becomes more pronounced. In particular, under $t_3$ and Cauchy errors, the bias and MISE of mean-based methods increases substantially, whereas OQRF remains stable, showing the robustness of the proposed quantile-based framework to heavy-tailed outcomes. Quantile-based competing methods, such as GRF and DML-QR, are more robust than mean-based methods, but they still perform less favorably than OQRF because they either do not adequately adjust for high-dimensional confounders or are less flexible in capturing heterogeneous treatment effects. The Monte Carlo standard errors reported in parentheses are small relative to the main performance gaps, indicating that the observed improvements are stable across replications.


We next evaluate the numerical performance of the little-bag inference procedure under Setting 1. For comparison, we also implement ORF and GRF using the same little-bag bootstrap scheme. As discussed in \cite{athey2019generalized}, valid inference for forest-based methods typically requires substantially more trees than are needed for point estimation alone. Accordingly, we set the number of trees to $B=4000$, use $G=50$ little bags with $R=80$ trees in each bag, and take the subsample size to be $s=0.3n$. Since this procedure is computationally expensive, we run 200 Monte Carlo replications. The results in Table~\ref{tab:inference_coverage} demonstrates that OQRF delivers the best overall performance, with reasonable confidence interval lengths and good empirical coverages. ORF deteriorates substantially under Cauchy errors, indicating high sensitivity to heavy-tailed noise. GRF also performs poorly, particularly in true value coverage rate, since it does not adjust for high-dimensional confounders,and may therefore yield substantially biased QTE estimates.

Additional numerical results are provided in Section S2 of the Online Supplementary Materials, including running time comparisons, small-sample, cross-sectional and model-misspecification studies and comparisons of CART and quantile-specific splitting strategies.


\section{Application to Real-World Non-Small Cell Lung Cancer Data}\label{sec:appl}

In patients with non-small cell lung cancer (NSCLC), levels of cell-free
circulating tumor DNA (ctDNA) have emerged as a promising biomarker for
monitoring disease burden and treatment response~\citep[see
e.g.,][]{sanz2022monitoring,bestvina2023early}.
Quantitative changes in ctDNA levels over time can reflect the underlying tumor
dynamics, often preceding radiographic evidence of response or progression~
\citep{anagnostou2023ctdna,
assaf2023longitudinal}.
Effective estimation of the individualized treatment effect through ctDNA
dynamics can therefore provide valuable insights into therapeutic efficacy,
particularly in the early stages of
intervention~\citep{goldberg2018early,ricciuti2021early}.

\begin{figure}[!p]
  \centering
  \subfigure[]{
    \includegraphics[width=0.46\textwidth]{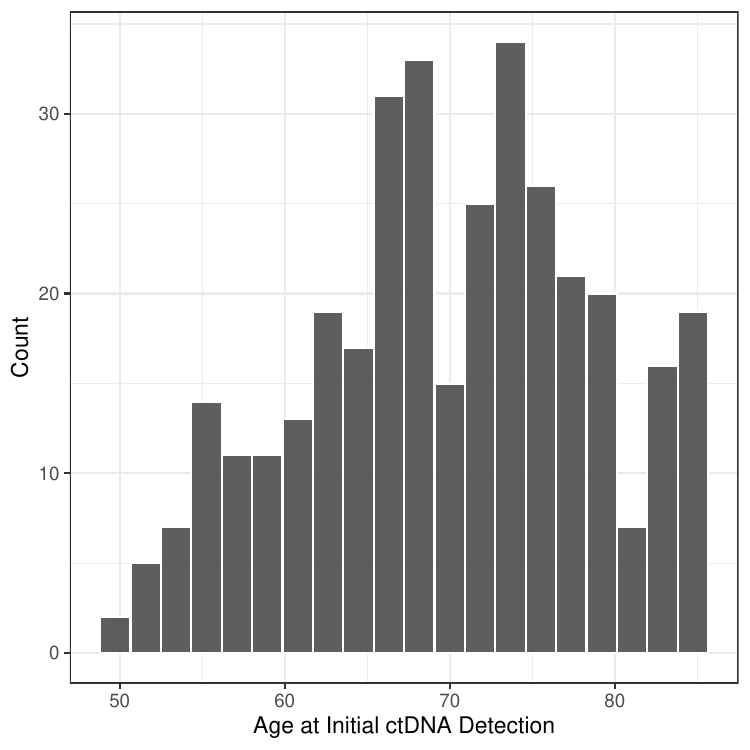}
    \label{fig:hist:age}
  }
  \subfigure[]{
    \includegraphics[width=0.46\textwidth]{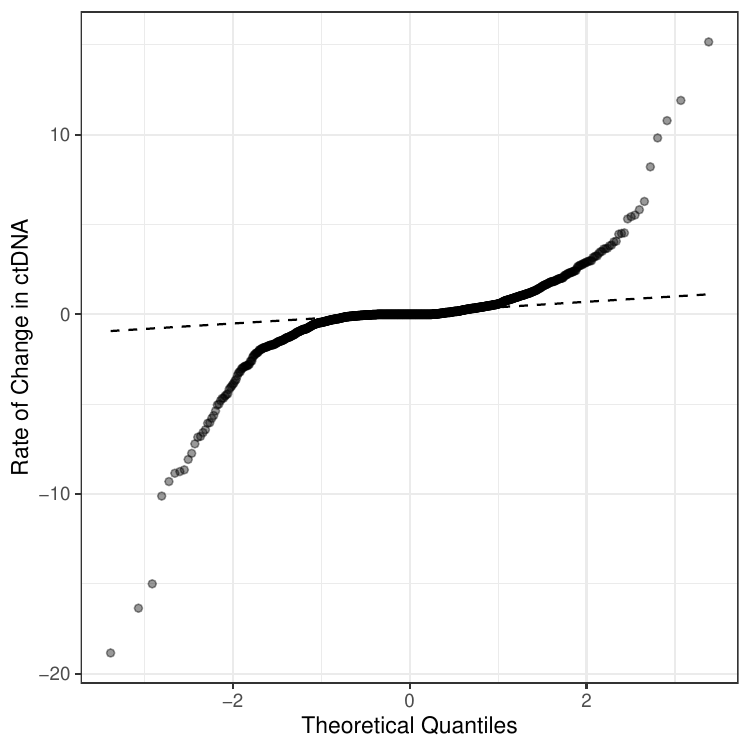}
    \label{fig:qq:ctdna}
  }
  \caption{(a) Histogram of patient age at the time of initial ctDNA detection
    and (b) quantile-quantile plot of the rate of change in
    ctDNA.}\label{fig:hist}
\end{figure}

We analyze a real-world dataset from the Flatiron Health Research
Database~\citep{flatiron2025data}, a database derived from the US electronic
health record that is deemed deidentified and comprised of over 280,000 patients
with NSCLC at the time of study.
Our study cohort included 346 patients diagnosed with NSCLC who were at least 50
years old and had serial ctDNA measurements collected during routine clinical
care between July 2017 and July 2024.
All patients had detectable ctDNA at baseline, and the dataset comprised 1,383
total observations of ctDNA change over time.
Our primary objective was to estimate heterogeneous treatment effects of
commonly used chemotherapy regimens, with exposure quantified as the
{\it time-averaged dose}, conditional on each patient’s age at initial ctDNA
detection.
The distribution of patient ages at first ctDNA detection is shown in
Figure~\ref{fig:hist:age}. To further assess the plausibility of the structural quantile model \eqref{eq:main_model} in this application, we provide a localized partial-residual diagnostic for the linearity-in-treatment assumption in Section~S3 of the Online Supplementary Materials.

In addition to demographic covariates, we took into account a total of 100
possible confounders, including the duration of NSCLC diagnosis, tumor stage,
advanced / metastatic disease status, previous surgery, smoking history,
molecular biomarkers, medication history other than chemotherapy, and
\emph{Eastern Cooperative Oncology Group} performance status.
The molecular biomarker data included both genomic alterations in FDA-approved
or emerging biomarkers (e.g., EGFR, ALK, ROS1, BRAF, MET, RET, NTRK1/2/3, KRAS,
ERBB2) and protein expression markers such as PD-L1.
For a recent overview of prognostic and predictive biomarkers in NSCLC, see
\citet{odintsov2024prognostic}.
As with all real-world EHR-derived data, residual confounding and informative
missingness cannot be fully excluded; see Section~\ref{sec:disc} for a
discussion of limitations.

We quantified the rate of change in ctDNA levels using the $\log_2$ fold change
with a pseudo-count of 1~\citep{erhard2018estimating} normalized by the time
interval in months.
As shown in Figure~\ref{fig:qq:ctdna}, the distribution of the rate of change in
ctDNA among NSCLC patients is markedly heavy-tailed compared to a normal
distribution, with a subset of individuals exhibiting extreme increases.
This heavy-tailed pattern likely reflects underlying biological heterogeneity,
differences in disease burden, and diverse treatment response dynamics.
These properties motivated our focus on estimating the conditional median
treatment effect rather than the mean, offering a more robust summary of the
typical patient response.

\begin{figure}[!p]
  \centering
  \includegraphics[width=0.8\textwidth]{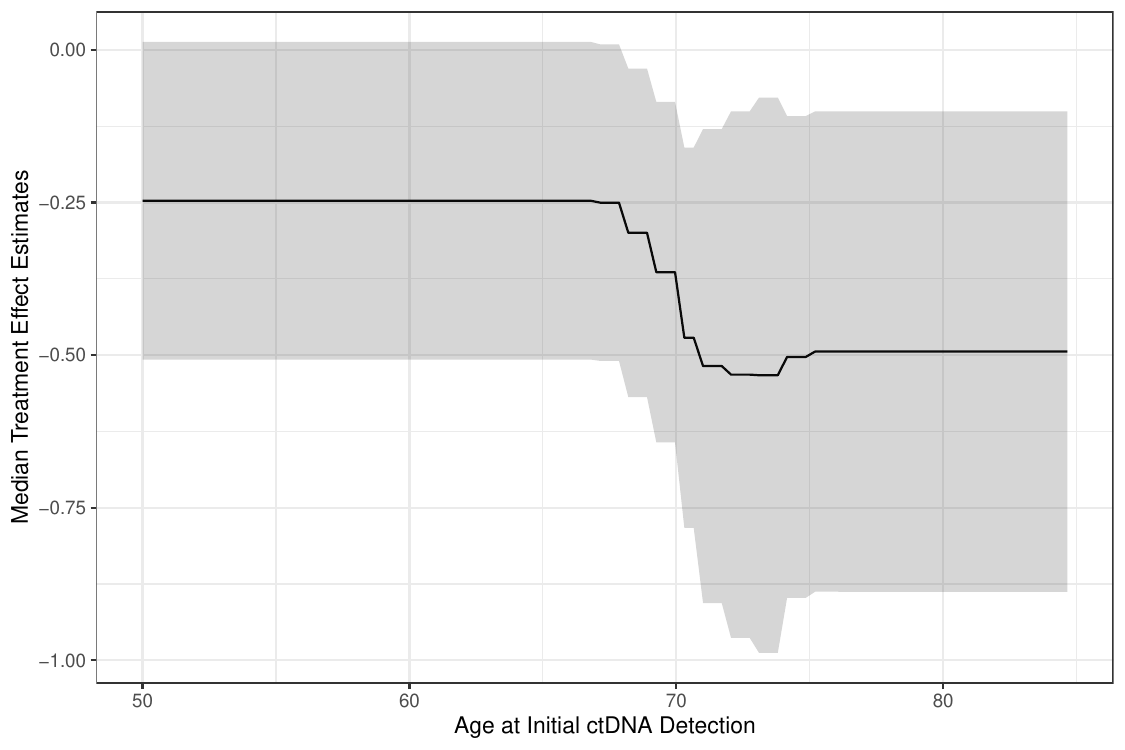}
  \caption{Median treatment effect estimates conditional on age at initial ctDNA
    detection, with 95\% pointwise confidence intervals indicated by shaded
    region.}\label{fig:theta-hat:50}
\end{figure}

We applied the proposed orthogonal quantile random forest method to estimate the
median treatment effect of chemotherapy, conditional on patient age at initial
ctDNA detection, as shown in Figure~\ref{fig:theta-hat:50}.
The pointwise 95\% confidence intervals constructed from the bag of little
bootstraps are indicated by shaded region.
The estimated treatment effects reveal a clear age-dependent trend.
Among patients younger than 67, the median treatment effect remained relatively
stable.
Between ages 67 and 73, chemotherapy appeared increasingly effective in reducing
ctDNA levels, with progressively more negative median effects.
Notably, for patients aged 74 or older, the median treatment effect diminished
slightly and plateaued rapidly after age 75, although it remained more favorable than in
the 50--70 age group.
While these results suggest a less pronounced benefit in older patients, they do
not indicate that age alone should preclude chemotherapy, which is consistent
with findings from prior
studies~\citep{weinmann2003treatment,veluswamy2016chemotherapy}.

In summary, our proposed approach enables interpretable, age-stratified
estimation of heterogeneous quantile treatment effects, with direct implications for personalizing chemotherapy decisions in NSCLC.

\section{Discussion}\label{sec:disc}
In this paper, we propose an OQRF framework to estimate heterogeneous quantile treatment effects in longitudinal studies with high-dimensional confounders and heavy-tailed outcomes. Our method has three main advantages, each of which is directly useful in the real data analysis. First, the orthogonal score makes the treatment effect estimator less sensitive to nuisance estimation errors, which is important in the NSCLC application, where many patient-level and clinical variables may act as confounders. Second, the quantile-specific splitting criterion and estimating equation are less sensitive to extreme outcomes than mean-based methods, which is useful to model the heavy-tailed outcomes in the NSCLC application. Third, the Bootstrap of Little Bags procedure provides practical uncertainty quantification, which helps assess the reliability of the estimated heterogeneous effects across patient characteristics. Together, these features make the proposed OQRF framework a robust and interpretable tool for studying heterogeneous treatment effects in complex biomedical longitudinal data.

Several limitations should be noted. First, the causal interpretation relies on the structural quantile model. This has two implications for the real data analysis. On one hand, the method adjusts for observed confounders, but it does not address hidden confounding. On the other hand, the model assumes that, after adjusting for the observed confounders, the conditional quantile of the outcome is locally linear in the current treatment. Therefore, when applying the method to real data, it is important to assess the plausibility of this linearity-in-treatment assumption. Second, the proposed estimand captures the effect of the current treatment on the current outcome quantile, rather than the effect of a full treatment trajectory \citep{li2024evaluating}; thus, dynamic-regime questions involving carryover effects or feedback mechanisms are beyond our scope. Finally, forest splitting is restricted to low-dimensional modifiers: high-dimensional variables are adjusted for as confounders but are not used to define treatment heterogeneity, so the real-data analysis explores nonlinear heterogeneity only over a limited set of baseline modifiers

In conclusion, the proposed OQRF framework provides a robust and flexible approach for estimating heterogeneous quantile treatment effects in longitudinal studies with high-dimensional confounding. At the same time, its causal interpretation relies on the structural quantile model and observed confounding adjustment, and the current framework focuses on contemporaneous treatment effects rather than full treatment trajectories. These limitations point to several important future directions.

\section*{Supplementary Materials}
Supplementary material is available online, including implementation details, additional simulation studies, model diagnostics in the real data analysis and proofs of the main theorems.


\section*{Data Availability statement}
The data underlying this study are derived from the Flatiron Health database and are not publicly available due to licensing restrictions.

\section*{Acknowledgments}
This work was partially supported by the National Key R$\&$D Program of China, [2023YFA1008700, 2023YFA1008703]; National Natural Science Foundation of China, [12371265]; Natural Science Foundation of Shanghai, [24ZR1455200]; Direct Grants for the Chinese University of Hong Kong, [171428926]; and Hong Kong Research Grants Council, [RGC-14308823].

\bibliographystyle{elsarticle-harv}
\bibliography{ref}

\begin{thebibliography}{34}
\expandafter\ifx\csname natexlab\endcsname\relax\def\natexlab#1{#1}\fi
\providecommand{\url}[1]{\texttt{#1}}
\providecommand{\href}[2]{#2}
\providecommand{\path}[1]{#1}
\providecommand{\DOIprefix}{doi:}
\providecommand{\ArXivprefix}{arXiv:}
\providecommand{\URLprefix}{URL: }
\providecommand{\Pubmedprefix}{pmid:}
\providecommand{\doi}[1]{\href{http://dx.doi.org/#1}{\path{#1}}}
\providecommand{\Pubmed}[1]{\href{pmid:#1}{\path{#1}}}
\providecommand{\bibinfo}[2]{#2}
\ifx\xfnm\relax \def\xfnm[#1]{\unskip,\space#1}\fi
\bibitem[{Anagnostou et~al.(2023)Anagnostou, Ho, Nicholas, Juergens, Sacher,
  Fung et~al.}]{anagnostou2023ctdna}
\bibinfo{author}{Anagnostou, V.}, \bibinfo{author}{Ho, C.},
  \bibinfo{author}{Nicholas, G.}, \bibinfo{author}{Juergens, R.A.},
  \bibinfo{author}{Sacher, A.}, \bibinfo{author}{Fung, A.S.}, et~al.,
  \bibinfo{year}{2023}.
\newblock \bibinfo{title}{{ctDNA} response after {P}embrolizumab in non-small
  cell lung cancer: {P}hase 2 adaptive trial results}.
\newblock \bibinfo{journal}{Nature Medicine} \bibinfo{volume}{29},
  \bibinfo{pages}{2559--2569}.
\bibitem[{Assaf et~al.(2023)Assaf, Zou, Fine, Socinski, Young, Lipson
  et~al.}]{assaf2023longitudinal}
\bibinfo{author}{Assaf, Z.J.F.}, \bibinfo{author}{Zou, W.},
  \bibinfo{author}{Fine, A.D.}, \bibinfo{author}{Socinski, M.A.},
  \bibinfo{author}{Young, A.}, \bibinfo{author}{Lipson, D.}, et~al.,
  \bibinfo{year}{2023}.
\newblock \bibinfo{title}{A longitudinal circulating tumor {DNA}-based model
  associated with survival in metastatic non-small-cell lung cancer}.
\newblock \bibinfo{journal}{Nature Medicine} \bibinfo{volume}{29},
  \bibinfo{pages}{859--868}.
\bibitem[{Athey et~al.(2019)Athey, Tibshirani and Wager}]{athey2019generalized}
\bibinfo{author}{Athey, S.}, \bibinfo{author}{Tibshirani, J.},
  \bibinfo{author}{Wager, S.}, \bibinfo{year}{2019}.
\newblock \bibinfo{title}{Generalized random forests}.
\newblock \bibinfo{journal}{The Annals of Statistics} \bibinfo{volume}{47},
  \bibinfo{pages}{1148--1178}.
\bibitem[{Baklicharov et~al.(2024)Baklicharov, Ley, Gorasso, Devleesschauwer
  and Vansteelandt}]{baklicharov2024assumption}
\bibinfo{author}{Baklicharov, G.}, \bibinfo{author}{Ley, C.},
  \bibinfo{author}{Gorasso, V.}, \bibinfo{author}{Devleesschauwer, B.},
  \bibinfo{author}{Vansteelandt, S.}, \bibinfo{year}{2024}.
\newblock \bibinfo{title}{Assumption-lean quantile regression}.
\newblock \bibinfo{journal}{arXiv preprint arXiv:2404.10495} .
\bibitem[{Belloni et~al.(2019)Belloni, Chernozhukov and
  Kato}]{belloni2019valid}
\bibinfo{author}{Belloni, A.}, \bibinfo{author}{Chernozhukov, V.},
  \bibinfo{author}{Kato, K.}, \bibinfo{year}{2019}.
\newblock \bibinfo{title}{Valid post-selection inference in high-dimensional
  approximately sparse quantile regression models}.
\newblock \bibinfo{journal}{Journal of the American Statistical Association}
  \bibinfo{volume}{114}, \bibinfo{pages}{749--758}.
\bibitem[{Bestvina et~al.(2023)Bestvina, Garassino, Neal, Wakelee, Diehn and
  Vokes}]{bestvina2023early}
\bibinfo{author}{Bestvina, C.M.}, \bibinfo{author}{Garassino, M.C.},
  \bibinfo{author}{Neal, J.W.}, \bibinfo{author}{Wakelee, H.A.},
  \bibinfo{author}{Diehn, M.}, \bibinfo{author}{Vokes, E.E.},
  \bibinfo{year}{2023}.
\newblock \bibinfo{title}{Early-stage lung cancer: {U}sing circulating tumor
  {DNA} to get personal}.
\newblock \bibinfo{journal}{Journal of Clinical Oncology} \bibinfo{volume}{41},
  \bibinfo{pages}{4093--4096}.
\bibitem[{Bica et~al.(2020)Bica, Jordon and van~der
  Schaar}]{bica2020estimating}
\bibinfo{author}{Bica, I.}, \bibinfo{author}{Jordon, J.},
  \bibinfo{author}{van~der Schaar, M.}, \bibinfo{year}{2020}.
\newblock \bibinfo{title}{Estimating the effects of continuous-valued
  interventions using generative adversarial networks}, in:
  \bibinfo{booktitle}{Advances in Neural Information Processing Systems}, pp.
  \bibinfo{pages}{16434--16445}.
\bibitem[{Chen et~al.(2025)Chen, Duan, Chernozhukov and
  Syrgkanis}]{chen2025automaticdoublyrobustforests}
\bibinfo{author}{Chen, Z.}, \bibinfo{author}{Duan, J.},
  \bibinfo{author}{Chernozhukov, V.}, \bibinfo{author}{Syrgkanis, V.},
  \bibinfo{year}{2025}.
\newblock \bibinfo{title}{Automatic doubly robust forests}.
\newblock \bibinfo{journal}{arXiv preprint arXiv:2412.07184} .
\bibitem[{Cheng et~al.(2024)Cheng, Hu and Li}]{cheng2024doubly}
\bibinfo{author}{Cheng, C.}, \bibinfo{author}{Hu, L.}, \bibinfo{author}{Li,
  F.}, \bibinfo{year}{2024}.
\newblock \bibinfo{title}{Doubly robust estimation and sensitivity analysis for
  marginal structural quantile models}.
\newblock \bibinfo{journal}{Biometrics} \bibinfo{volume}{80},
  \bibinfo{pages}{ujae045}.
\bibitem[{Chernozhukov et~al.(2018)Chernozhukov, Chetverikov, Demirer, Duflo,
  Hansen, Newey et~al.}]{chernozhukov2018double}
\bibinfo{author}{Chernozhukov, V.}, \bibinfo{author}{Chetverikov, D.},
  \bibinfo{author}{Demirer, M.}, \bibinfo{author}{Duflo, E.},
  \bibinfo{author}{Hansen, C.}, \bibinfo{author}{Newey, W.}, et~al.,
  \bibinfo{year}{2018}.
\newblock \bibinfo{title}{{Double/Debiased Machine Learning for Treatment and
  Structural Parameters}}.
\newblock \bibinfo{journal}{The Econometrics Journal} \bibinfo{volume}{21},
  \bibinfo{pages}{1--68}.
\bibitem[{Datta and Beck(2014)}]{datta2014robust}
\bibinfo{author}{Datta, S.}, \bibinfo{author}{Beck, J.D.},
  \bibinfo{year}{2014}.
\newblock \bibinfo{title}{Robust estimation of marginal regression parameters
  in clustered data}.
\newblock \bibinfo{journal}{Statistical modelling} \bibinfo{volume}{14},
  \bibinfo{pages}{489--501}.
\bibitem[{Doss et~al.(2024)Doss, Weng, Wang, Moscovice and
  Chantarat}]{doss2024nonparametric}
\bibinfo{author}{Doss, C.R.}, \bibinfo{author}{Weng, G.},
  \bibinfo{author}{Wang, L.}, \bibinfo{author}{Moscovice, I.},
  \bibinfo{author}{Chantarat, T.}, \bibinfo{year}{2024}.
\newblock \bibinfo{title}{A nonparametric doubly robust test for a continuous
  treatment effect}.
\newblock \bibinfo{journal}{The Annals of Statistics} \bibinfo{volume}{52},
  \bibinfo{pages}{1592--1615}.
\bibitem[{Emmenegger and B{\"u}hlmann(2023)}]{emmenegger2023plug}
\bibinfo{author}{Emmenegger, C.}, \bibinfo{author}{B{\"u}hlmann, P.},
  \bibinfo{year}{2023}.
\newblock \bibinfo{title}{Plug-in machine learning for partially linear
  mixed-effects models with repeated measurements}.
\newblock \bibinfo{journal}{Scandinavian Journal of Statistics}
  \bibinfo{volume}{50}, \bibinfo{pages}{1553--1567}.
\bibitem[{Erhard(2018)}]{erhard2018estimating}
\bibinfo{author}{Erhard, F.}, \bibinfo{year}{2018}.
\newblock \bibinfo{title}{Estimating pseudocounts and fold changes for digital
  expression measurements}.
\newblock \bibinfo{journal}{Bioinformatics} \bibinfo{volume}{34},
  \bibinfo{pages}{4054--4063}.
\bibitem[{Fernandes et~al.(2021)Fernandes, Guerre and
  Horta}]{fernandes2021smoothing}
\bibinfo{author}{Fernandes, M.}, \bibinfo{author}{Guerre, E.},
  \bibinfo{author}{Horta, E.}, \bibinfo{year}{2021}.
\newblock \bibinfo{title}{Smoothing quantile regressions}.
\newblock \bibinfo{journal}{Journal of Business \& Economic Statistics}
  \bibinfo{volume}{39}, \bibinfo{pages}{338--357}.
\bibitem[{{Flatiron Health}(2025)}]{flatiron2025data}
\bibinfo{author}{{Flatiron Health}}, \bibinfo{year}{2025}.
\newblock \bibinfo{title}{Database characterization guide}.
\newblock
  \bibinfo{howpublished}{\url{https://flatiron.com/database-characterization}}.
\newblock \bibinfo{note}{Published March 18, 2025. Accessed August 3, 2025}.
\bibitem[{Foster and Syrgkanis(2023)}]{foster2023orthogonal}
\bibinfo{author}{Foster, D.J.}, \bibinfo{author}{Syrgkanis, V.},
  \bibinfo{year}{2023}.
\newblock \bibinfo{title}{Orthogonal statistical learning}.
\newblock \bibinfo{journal}{The Annals of Statistics} \bibinfo{volume}{51},
  \bibinfo{pages}{879--908}.
\bibitem[{Fr{\"o}lich and Melly(2013)}]{frolich2013unconditional}
\bibinfo{author}{Fr{\"o}lich, M.}, \bibinfo{author}{Melly, B.},
  \bibinfo{year}{2013}.
\newblock \bibinfo{title}{Unconditional quantile treatment effects under
  endogeneity}.
\newblock \bibinfo{journal}{Journal of Business \& Economic Statistics}
  \bibinfo{volume}{31}, \bibinfo{pages}{346--357}.
\bibitem[{Fuhr and Papies(2024)}]{fuhr2024double}
\bibinfo{author}{Fuhr, J.}, \bibinfo{author}{Papies, D.}, \bibinfo{year}{2024}.
\newblock \bibinfo{title}{Double machine learning meets panel data--promises,
  pitfalls, and potential solutions}.
\newblock \bibinfo{journal}{arXiv preprint arXiv:2409.01266} .
\bibitem[{Giessing and Wang(2023)}]{giessing2023debiased}
\bibinfo{author}{Giessing, A.}, \bibinfo{author}{Wang, J.},
  \bibinfo{year}{2023}.
\newblock \bibinfo{title}{Debiased inference on heterogeneous quantile
  treatment effects with regression rank scores}.
\newblock \bibinfo{journal}{Journal of the Royal Statistical Society Series B:
  Statistical Methodology} \bibinfo{volume}{85}, \bibinfo{pages}{1561--1588}.
\bibitem[{Goldberg et~al.(2018)Goldberg, Narayan, Kole, Decker, Teysir,
  Carriero et~al.}]{goldberg2018early}
\bibinfo{author}{Goldberg, S.B.}, \bibinfo{author}{Narayan, A.},
  \bibinfo{author}{Kole, A.J.}, \bibinfo{author}{Decker, R.H.},
  \bibinfo{author}{Teysir, J.}, \bibinfo{author}{Carriero, N.J.}, et~al.,
  \bibinfo{year}{2018}.
\newblock \bibinfo{title}{Early assessment of lung cancer immunotherapy
  response via circulating tumor {DNA}}.
\newblock \bibinfo{journal}{Clinical Cancer Research} \bibinfo{volume}{24},
  \bibinfo{pages}{1872--1880}.
\bibitem[{Kennedy et~al.(2017)Kennedy, Ma, McHugh and Small}]{kennedy2017non}
\bibinfo{author}{Kennedy, E.H.}, \bibinfo{author}{Ma, Z.},
  \bibinfo{author}{McHugh, M.D.}, \bibinfo{author}{Small, D.S.},
  \bibinfo{year}{2017}.
\newblock \bibinfo{title}{Non-parametric methods for doubly robust estimation
  of continuous treatment effects}.
\newblock \bibinfo{journal}{Journal of the Royal Statistical Society Series B:
  Statistical Methodology} \bibinfo{volume}{79}, \bibinfo{pages}{1229--1245}.
\bibitem[{K{\"u}nzel et~al.(2019)K{\"u}nzel, Sekhon, Bickel and
  Yu}]{kunzel2019metalearners}
\bibinfo{author}{K{\"u}nzel, S.R.}, \bibinfo{author}{Sekhon, J.S.},
  \bibinfo{author}{Bickel, P.J.}, \bibinfo{author}{Yu, B.},
  \bibinfo{year}{2019}.
\newblock \bibinfo{title}{Metalearners for estimating heterogeneous treatment
  effects using machine learning}.
\newblock \bibinfo{journal}{Proceedings of the national academy of sciences}
  \bibinfo{volume}{116}, \bibinfo{pages}{4156--4165}.
\bibitem[{Li et~al.(2024)Li, Shi, Lu, Li and Zhu}]{li2024evaluating}
\bibinfo{author}{Li, T.}, \bibinfo{author}{Shi, C.}, \bibinfo{author}{Lu, Z.},
  \bibinfo{author}{Li, Y.}, \bibinfo{author}{Zhu, H.}, \bibinfo{year}{2024}.
\newblock \bibinfo{title}{Evaluating dynamic conditional quantile treatment
  effects with applications in ridesharing}.
\newblock \bibinfo{journal}{Journal of the American Statistical Association}
  \bibinfo{volume}{119}, \bibinfo{pages}{1736--1750}.
\bibitem[{Odintsov and Sholl(2024)}]{odintsov2024prognostic}
\bibinfo{author}{Odintsov, I.}, \bibinfo{author}{Sholl, L.M.},
  \bibinfo{year}{2024}.
\newblock \bibinfo{title}{Prognostic and predictive biomarkers in non-small
  cell lung carcinoma}.
\newblock \bibinfo{journal}{Pathology} \bibinfo{volume}{56},
  \bibinfo{pages}{192--204}.
\bibitem[{Oprescu et~al.(2019)Oprescu, Syrgkanis and Wu}]{pmlr-v97-oprescu19a}
\bibinfo{author}{Oprescu, M.}, \bibinfo{author}{Syrgkanis, V.},
  \bibinfo{author}{Wu, Z.S.}, \bibinfo{year}{2019}.
\newblock \bibinfo{title}{Orthogonal random forest for causal inference}, in:
  \bibinfo{booktitle}{International Conference on Machine Learning},
  \bibinfo{organization}{PMLR}. p. \bibinfo{pages}{4932–4941}.
\bibitem[{Qu and Li(2006)}]{qu2006quadratic}
\bibinfo{author}{Qu, A.}, \bibinfo{author}{Li, R.}, \bibinfo{year}{2006}.
\newblock \bibinfo{title}{Quadratic inference functions for varying-coefficient
  models with longitudinal data}.
\newblock \bibinfo{journal}{Biometrics} \bibinfo{volume}{62},
  \bibinfo{pages}{379--391}.
\bibitem[{Ricciuti et~al.(2021)Ricciuti, Jones, Severgnini, Alessi, Recondo,
  Lawrence et~al.}]{ricciuti2021early}
\bibinfo{author}{Ricciuti, B.}, \bibinfo{author}{Jones, G.},
  \bibinfo{author}{Severgnini, M.}, \bibinfo{author}{Alessi, J.V.},
  \bibinfo{author}{Recondo, G.}, \bibinfo{author}{Lawrence, M.}, et~al.,
  \bibinfo{year}{2021}.
\newblock \bibinfo{title}{Early plasma circulating tumor {DNA} (ctdna) changes
  predict response to first-line pembrolizumab-based therapy in non-small cell
  lung cancer (nsclc)}.
\newblock \bibinfo{journal}{Journal for Immunotherapy of Cancer}
  \bibinfo{volume}{9}, \bibinfo{pages}{e001504}.
\bibitem[{Sanz-Garcia et~al.(2022)Sanz-Garcia, Zhao, Bratman and
  Siu}]{sanz2022monitoring}
\bibinfo{author}{Sanz-Garcia, E.}, \bibinfo{author}{Zhao, E.},
  \bibinfo{author}{Bratman, S.V.}, \bibinfo{author}{Siu, L.L.},
  \bibinfo{year}{2022}.
\newblock \bibinfo{title}{Monitoring and adapting cancer treatment using
  circulating tumor {DNA} kinetics: {C}urrent research, opportunities, and
  challenges}.
\newblock \bibinfo{journal}{Science Advances} \bibinfo{volume}{8},
  \bibinfo{pages}{eabi8618}.
\bibitem[{Tan et~al.(2022)Tan, Wang and Zhou}]{tan2022high}
\bibinfo{author}{Tan, K.M.}, \bibinfo{author}{Wang, L.}, \bibinfo{author}{Zhou,
  W.X.}, \bibinfo{year}{2022}.
\newblock \bibinfo{title}{High-dimensional quantile regression: Convolution
  smoothing and concave regularization}.
\newblock \bibinfo{journal}{Journal of the Royal Statistical Society Series B:
  Statistical Methodology} \bibinfo{volume}{84}, \bibinfo{pages}{205--233}.
\bibitem[{Veluswamy et~al.(2016)Veluswamy, Levy and
  Wisnivesky}]{veluswamy2016chemotherapy}
\bibinfo{author}{Veluswamy, R.R.}, \bibinfo{author}{Levy, B.},
  \bibinfo{author}{Wisnivesky, J.P.}, \bibinfo{year}{2016}.
\newblock \bibinfo{title}{Chemotherapy in elderly patients with nonsmall cell
  lung cancer}.
\newblock \bibinfo{journal}{Current Opinion in Pulmonary Medicine}
  \bibinfo{volume}{22}, \bibinfo{pages}{336--343}.
\bibitem[{Wager and Athey(2018)}]{wager2018estimation}
\bibinfo{author}{Wager, S.}, \bibinfo{author}{Athey, S.}, \bibinfo{year}{2018}.
\newblock \bibinfo{title}{Estimation and inference of heterogeneous treatment
  effects using random forests}.
\newblock \bibinfo{journal}{Journal of the American Statistical Association}
  \bibinfo{volume}{113}, \bibinfo{pages}{1228--1242}.
\bibitem[{Weinmann et~al.(2003)Weinmann, Jeremic, Toomes, Friedel and
  Bamberg}]{weinmann2003treatment}
\bibinfo{author}{Weinmann, M.}, \bibinfo{author}{Jeremic, B.},
  \bibinfo{author}{Toomes, H.}, \bibinfo{author}{Friedel, G.},
  \bibinfo{author}{Bamberg, M.}, \bibinfo{year}{2003}.
\newblock \bibinfo{title}{Treatment of lung cancer in the elderly. part i:
  {N}on-small cell lung cancer}.
\newblock \bibinfo{journal}{Lung Cancer} \bibinfo{volume}{39},
  \bibinfo{pages}{233--253}.
\bibitem[{Zhu et~al.(2022)Zhu, Sun and Wei}]{zhu2022hybrid}
\bibinfo{author}{Zhu, H.}, \bibinfo{author}{Sun, Y.}, \bibinfo{author}{Wei,
  Y.}, \bibinfo{year}{2022}.
\newblock \bibinfo{title}{Hybrid censored quantile regression forest to assess
  the heterogeneous effects}.
\newblock \bibinfo{journal}{arXiv preprint arXiv:2212.05672} .

\end{thebibliography}

\begin{sidewaystable}
\centering
\caption{Bias and MISE comparison across methods when sample size $2n=2000$}
\label{tab:1}
\renewcommand{\arraystretch}{1.5}
\resizebox{\linewidth}{!}{
\begin{tabular}{ll|cccccc|cccccc}
\toprule
\multicolumn{2}{c}{}&\multicolumn{6}{|c|}{\textbf{Bias}($*10^{-2}$)}& \multicolumn{6}{c}{\textbf{MISE}($*10^{-2}$)}\\
\midrule
&  & OQRF  & ORF & GRF & DML-Lasso & DML-RF& DML-QR & OQRF  & ORF &GRF& DML-Lasso & DML-RF& DML-QR \\
\midrule
\multicolumn{2}{c|}{} & \multicolumn{12}{c}{Setting 1}\\
\midrule
\multirow{3}{*}{$p_w=201$} 
& Normal & $\bm{0.22}$(0.11)  & 0.80(0.12) & -14.06(0.05) & 0.31(0.09) & 13.41(0.07) & 18.52(0.24) & $\bm{0.49}$(0.01)  & 0.61(0.01) & 2.22(0.22) & 0.90(0.01) & 2.83(0.02) & 5.84(0.11)\\
& $t_3$   & $\bm{0.27}$(0.12)  & 1.70(0.22) & -16.04(0.06) & 0.31(0.15) & 13.45(0.10) & 18.07(0.25) &  $\bm{0.59}$(0.01)  & 1.20(0.03) & 2.88(0.02) & 1.21(0.02) & 2.97(0.03) & 5.71(0.11)\\
& Cauchy & $\bm{1.04}$(0.15)  & 10.74(40.91) & -19.86(0.07) & -5.91(29.34) & 99.37($1.03*10^2$) & 18.74(0.33) & $\bm{0.84}$(0.02)  & $3.44*10^4$($1.31*10^4$) & 4.38(0.03) & $1.70*10^5$($4.74*10^3$) & $2.09*10^5$($1.94*10^5$) & 6.83(0.16) \\
\midrule

\multirow{3}{*}{$p_w=501$} 
& Normal & 0.50(0.10)  & 1.14(0.13) & -15.24(0.05) & $\bm{0.25}$(0.08) & 13.02(0.06) & 22.92(0.26) & $\bm{0.48}$(0.01)  & 0.78(0.01) & 2.58(0.02) & 0.94(0.01) & 2.65(0.02) & 7.68(0.14) \\
& $t_3$   & $\bm{0.81}$(0.12)  & 2.59(0.21)  & -17.28(0.06) & 0.88(0.15) & 13.16(0.08)& 22.54(0.27) & $\bm{0.58}$(0.01)  & 1.31(0.04) & 3.29(0.02) & 1.28(0.02) & 2.77(0.02) & 7.69(0.14)\\
& Cauchy  & $\bm{1.38}$(0.15)  & 47.94(49.98) & -21.57(0.08) & 34.10(72.16) & 4.74(40.47) & 22.65(0.33) & $\bm{0.83}$(0.02)  & $3.57*10^4$($1.23*10^4$) & 5.11(0.03) &$1.52*10^5$($8.88*10^4$) & $3.49*10^4$($2.02*10^4$) & 8.44(0.17)\\
\midrule

\multicolumn{2}{c|}{} & \multicolumn{12}{c}{Setting 2}\\
\midrule
\multirow{3}{*}{$p_w=201$} 
& Normal  &  0.35(0.11) & 0.30(0.12) & -19.10(0.06) & $\bm{0.18}$(0.09) & 9.70(0.07)& 6.37(0.24) &  $\bm{0.56}$(0.01) & 0.67(0.01) & 4.14(0.03) & 0.85(0.01) & 1.90(0.02) & 6.93(0.10)\\
& $t_3$   & $\bm{0.34}$(0.13) & 1.21(0.20) & -21.68(0.07) & 0.51(0.14) & 9.80(0.10) & 6.36(0.26) & $\bm{0.67}$(0.01) & 1.19(0.03) & 5.29(0.03) & 1.14(0.02) & 2.07(0.02) & 6.94(0.11)\\
& Cauchy  & $\bm{1.03}(0.16)$ & 2.07(39.61) & -26.95(0.10) & -31.78(28.65) & $1.44*10^2$($1.28*10^2$) & 6.55(0.30)   &  $\bm{0.95}(0.02)$ & $3.18* 10^4$($1.23*10^4$) & 8.18(0.06) & $1.54* 10^4$($3.79*10^3$) & $2.99*10^5$($2.89*10^5$) & 7.68(0.14) \\
\midrule

\multirow{3}{*}{$p_w=501$} 
& Normal  & 0.70(0.11) & $\bm{0.42}$(0.12) & -20.56(0.06) & 0.68(0.08) & 9.51(0.07) & 14.32(0.18) & $\bm{0.56}$(0.01) & 0.88(0.01) & 4.73(0.03) & -0.85($4.71*10^{-3}$) & 1.82(0.01) & 8.70(0.11) \\
& $t_3$  & $\bm{1.01}(0.13)$ & 2.02(0.19) & -23.29(0.07) & 2.29(0.14) & 9.65(0.11) & 14.50(0.21) & $\bm{0.66}$(0.01) & 1.37(0.03) & 6.06(0.03) & 1.20(0.02) & 1.96(0.02) & 9.01(0.14) \\
& Cauchy & $\bm{1.37}$(0.16) & 29.29(46.85) & -29.26(0.10) & 36.68(64.04) & 29.83(45.60) & 14.18(0.24) & $\bm{0.97}$(0.02) & $2.56*10^4$($9.09*10^3$) & 9.57(0.06) & $9.05* 10^4$($3.86*10^4$) & $3.51* 10^4$($1.45* 10^4$) & 9.42(0.16)\\
\midrule
\multicolumn{2}{c|}{} & \multicolumn{12}{c}{Setting 3}\\
\midrule
\multirow{3}{*}{$p_w=201$} 
& Normal  & 0.58(0.11)  & 0.78(0.14) & -14.35(0.05) & $\bm{0.21}$(0.09) & 13.98(0.07) & 20.89(0.27) &  $\bm{1.00}$(0.02) & 6.36(0.05) & 2.88(0.02) & 1.54(0.01) & 4.19(0.03) & 9.57(0.16)\\
& $t_3$   & $\bm{0.72}$(0.13) & 1.59(0.23) &-16.14(0.06) & 0.79(0.15)  & 13.96(0.10) & 21.46(0.30) &  $\bm{1.15}$(0.02) & 8.09(0.07) & 3.63(0.02) & 2.06(0.03) & 4.40(0.03) & 10.18(0.19)\\
& Cauchy  &  $\bm{1.47}$(0.17)  & 41.06(70.82) & -20.15(0.08) & -42.31($1.18*10^2$) & 73.66(24.17) & 32.58(0.33)& $\bm{1.71}$(0.03) & $8.25*10^4$($4.22*10^4$) & 5.76(0.04) & $3.58*10^5$($2.25*10^5$) & $3.94*10^4$($2.02*10^4$) & 11.41(0.21)\\
\midrule

\multirow{3}{*}{$p_w=501$} 
& Normal  & 1.01(0.11)  & 1.25(0.15) & -15.72(0.05) & $\bm{0.23}$(0.07) & 13.78(0.07) & 25.45(0.29) & $\bm{0.96}$(0.01) & 7.67(0.05) & 4.10(0.02) & 1.61(0.01) & 3.83(0.02) & 12.06(0.20)\\
& $t_3$ & $\bm{1.38}$(0.13)  & 2.41(0.02) & -18.21(0.06) &1.51(0.15) & 14.14(0.09) & 25.27(0.29) & $\bm{1.18}$(0.02)  & 9.19(0.08) & 6.07(0.04) & 2.21(0.04) & 4.01(0.03) & 12.22(0.21)\\
& Cauchy  & $\bm{1.77}$(0.16) & 3.72(87.23) & -23.93(0.09) & -58.15(78.02) & 65.09(21.28) & 24.97(0.35) & $\bm{1.63}$(0.03)  & $9.07*10^4$($6.75*10^4$) & 11.68(0.08) & $1.75*10^5$($1.03*10^5$) & $1.47*10^4$($4.83*10^3$) & 13.39(0.26)\\
\bottomrule

\end{tabular}}

\vspace{0.5em} \begin{minipage}{\linewidth} \small
\textit{Note.} Each entry is reported as the Monte Carlo estimate with its standard error in parentheses. Bold numbers indicate the best performance among all methods. For the $r$th replication, define $\mathrm{Bias}^{(r)} = \frac{1}{100} \sum_{j=1}^{100} \left\{ \hat{\theta}^{(r)}(x_j)-\theta(x_j) \right\}$ and $\mathrm{ISE}^{(r)} = \frac{1}{100} \sum_{j=1}^{100} \left\{ \hat{\theta}^{(r)}(x_j)-\theta(x_j) \right\}^{2}.$ The reported Bias, MISE and their standard errors are computed as
$\mathrm{Bias} = \frac{1}{500} \sum_{r=1}^{500} \mathrm{Bias}^{(r)}$, $\mathrm{MISE} = \frac{1}{500} \sum_{r=1}^{500} \mathrm{ISE}^{(r)},\quad\mathrm{SE}(\mathrm{Bias}) = \frac{ \mathrm{sd}\left( \mathrm{Bias}^{(1)},\ldots,\mathrm{Bias}^{(500)} \right) }{\sqrt{500}}, \qquad \mathrm{SE}(\mathrm{MISE}) = \frac{ \mathrm{sd}\left( \mathrm{ISE}^{(1)},\ldots,\mathrm{ISE}^{(500)} \right) }{\sqrt{500}}$.
\end{minipage}
\end{sidewaystable}

\begin{table}[!p]
\centering
\caption{Empirical coverage and average confidence interval length under different error distributions. 
The columns ``True value coverage'' report the empirical coverage of confidence intervals for the true target parameter, while the columns ``Expected prediction coverage'' report the empirical coverage for the expected forest prediction, defined as the Monte Carlo average of the forest point estimates over 200 replications. This quantity is commonly used to assess whether the variance estimator captures the sampling variability of the forest prediction itself \citep{athey2019generalized}. The column ``CI Length'' report the average confidence interval length.}
\label{tab:inference_coverage}
\renewcommand{\arraystretch}{1.5}
\resizebox{\textwidth}{!}{
\begin{tabular}{llcccccccc}
\toprule
\multirow{2}{*}{ } 
& \multirow{2}{*}{Method}
& \multicolumn{2}{c}{True value coverage}
& \multicolumn{2}{c}{Expected prediction coverage}
& \multicolumn{2}{c}{CI length} \\
\cmidrule(lr){3-4}
\cmidrule(lr){5-6}
\cmidrule(lr){7-8}
& 
& 95\% CI & 90\% CI
& 95\% CI & 90\% CI
& 95\% CI & 90\% CI \\
\midrule

\multirow{3}{*}{Normal}
& OQRF 
& 0.943 & 0.896 
& 0.946 & 0.901 
& 0.325 & 0.273 \\
& ORF
& 0.911 & 0.851 
& 0.925 & 0.869 
& 0.340 & 0.286 \\
& GRF
& 0.044 & 0.021 
& 0.883 & 0.817
& 0.195 & 0.164 \\

\midrule

\multirow{3}{*}{Cauchy}
& OQRF
& 0.962 & 0.923 
& 0.965 & 0.929 
& 0.513 & 0.430  \\
& ORF 
& 0.917 & 0.856 
& 0.784 & 0.716 
& 16.890 & 14.174  \\
& GRF 
& 0.070 & 0.033 
& 0.888 & 0.826 
& 0.334 & 0.281  \\

\bottomrule
\end{tabular}
}
\end{table}

\end{document}